\documentclass{ifacconf}
\usepackage{wrapfig}
\renewcommand{\thefootnote}{\fnsymbol{footnote}}
\renewcommand{\thefootnote}{\fnsymbol{footnote}}
\usepackage{amsfonts}
\usepackage{caption}
\usepackage{enumitem}
\usepackage{graphicx}      
\usepackage{makecell}
\usepackage{caption}
\usepackage{natbib}        
\usepackage{xcolor}
\usepackage{booktabs}
\usepackage{amsmath}
\usepackage{subcaption}
\usepackage{multirow}
\usepackage{bm}

\renewcommand{\thefootnote}{\fnsymbol{footnote}}
\newcommand{\Transp}{\mathsf{T}}
\begin{document}
\begin{frontmatter}

\title{Physics-Guided Recurrent State-Space Neural Networks for Multi-Step Prediction } 

\thanks[footnoteinfo]{This work is funded by the Sensor AI Lab, under the AI Labs program of Delft University of Technology, and by the Convergence Human Mobility Center.}

\author[First]{Ruiyuan Li} 
\author[Second]{Ajay Seth} 
\author[First]{Manon Kok}

\address[First]{Delft Center for Systems and Control, TU Delft, the Netherlands,
       email: { \{r.li-6,m.kok-1\}@tudelft.nl}.}
\address[Second]{Department of Biomechanical Engineering, TU Delft, the Netherlands,
       email:  a.seth@tudelft.nl.}

\begin{abstract}                
State-space models are traditionally based on physical knowledge, but multi-step predictions from these physical models can be poor due to model inaccuracy. Black-box deep learning has shown promise as an alternative. However, these methods rely on the availability of large datasets and potentially available physical knowledge is neglected.  We propose the PG-RSSNN, a physics-guided recurrent state-space neural network that incorporates recurrent structures to enable the use of non-saturating activation functions in multi-step prediction. It mitigates the vanishing gradients and eliminates the risk of numerical divergence in training seen in existing structures that feed back state estimates. Results across multiple systems with various physical model imperfections, from linear state-space models with Gaussian noise to a robotic arm and a cascaded water tank system, show that the proposed PG-RSSNN maintains stable training behavior, and improves multi-step predictions, as compared with black-box neural networks and physics-only models, even with limited training data and when physical models are only partially known.
\end{abstract} 

\begin{keyword}
State-space models, physics-guided learning, recurrent neural networks, nonlinear systems, cascaded water tank system.
\end{keyword}
\end{frontmatter}

\section{Introduction}
\label{sec:introduction}
State-space models (SSMs) are commonly used for state estimation, e.g.\ in the fields of control and sensor fusion~\citep{aastrom2021feedback, gustafsson2010statistical}, and are typically based on physical knowledge. However, in practice, models are often inaccurate or only partially known 
because physical parameters and complete system inputs are difficult or impossible to measure \citep{verhaegen_filtering_2007,yu2020identification}. Because of this, \citet{suykens1995nonlinear} proposed a state-space neural network (SSNN). This network mirrors the state-space structure to learn nonlinear systems in a black-box manner. Recently, the potential of using deeper neural networks (NNs) to model nonlinear dynamical systems has also been demonstrated in \citet{gedon_deep_2021}.  However, black-box learning typically requires large amounts of training data \citep{schoukens2020cascaded} and potentially available physical knowledge is neglected. Thus, in recent years, physics-guided neural networks (PGNNs) \citep{daw_physics-guided_2021} have been developed to leverage existing physical knowledge while simultaneously learning the unknown parts of the model using NNs. Inspired by this, \citet{liu_physics-guided_2024} proposed W-PGNN, a physics-guided SSNN for nonlinear SSM identification.

Motivated by these works, we propose a Physics-Guided Recurrent State-Space Neural Network  (PG-RSSNN), which mainly differs from  \citet{liu_physics-guided_2024} in two ways: 1) our model uses a recurrent structure instead of directly feeding back the states to the NNs, and 2) it uses the non-saturating rectified linear unit (ReLU) activation function, requiring no dataset-specific parameter
tuning, to mitigate the widely known issue of vanishing gradients \citep{glorot2011deep}. 
The RNN structure avoids potential exponential growth of the state, which can occur in multi-step prediction when feeding the state back with non-saturating activations, as done e.g.\ in \citet{liu_physics-guided_2024}.

We demonstrate that our proposed PG-RSSNN maintains stable training behavior and achieves reduced multi-step prediction error as compared to black-box Recurrent State-Space Neural Networks (RSSNNs), physics-only models and other architectural variants across diverse systems: a linear state-space model with Gaussian noise, a real-world industrial robotic arm \citep{weigand_dataset_nodate}, a simulated robotic arm, and a cascaded water tank system \citep{schoukens2020cascaded}.  
The proposed model can be straightforwardly adapted to different systems with only minor modifications  limited to adjusting the input, state, output, and hidden layer dimensions of the NNs. 

\begin{figure*}[t]
\centering
\includegraphics[width=0.9\textwidth, trim=0.1cm 1.4cm 0.6cm 0cm, clip]{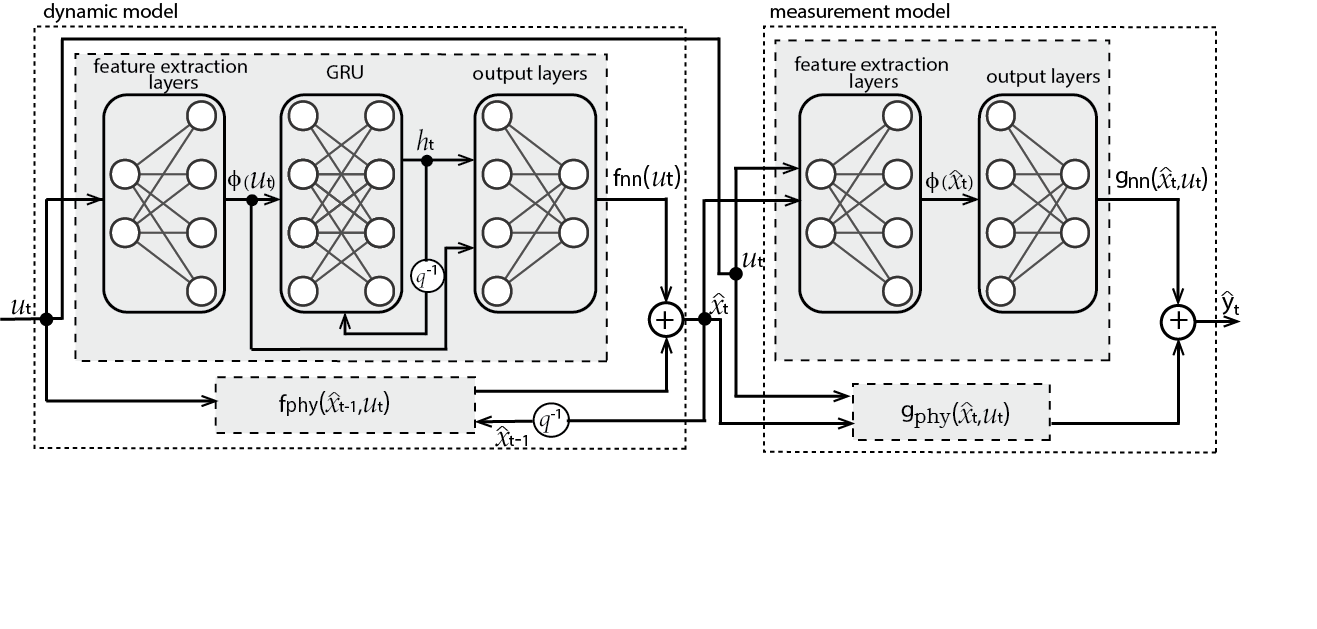}
\caption{The proposed PG-RSSNN: we adapt the GRU layer, and the estimated state $\hat{x}_t$ is not fed back to $f_{\text{nn}}(\cdot)$.}\label{fig:PG-RSSNN}
\end{figure*}
\section{Related work}

Driven by the aim of combining data-driven NNs with physical models for nonlinear system identification, several approaches have been proposed. Existing work has either adapted the model structure \citep{relan2017unstructured}, or the loss function, or both \citep{liu_physics-guided_2024, schon2022multi}. Our focus is on the model structure.  

One established approach that adapts the model structure to combine data-driven NNs with physical models is NLSS2 \citep{relan2017unstructured}.
Their approach learns the model by splitting it into a purely physics-based linear model fitted using a best linear approximation, and a purely data-driven model to learn the remaining nonlinearities. To reduce sensitivity to accurate state estimates, especially for multi-step predictions, we instead choose a tighter integration of the physics-based and data-driven models. This is inspired by \citet{schon2022multi} and \citet{liu_physics-guided_2024}, who adapt end-to-end physics-guided NNs, both feeding the estimated state to the NNs to learn the dynamics, similar to the data flow in an SSM. \citet{schon2022multi} adapt the model structure, feeding the output of the physical model to NNs, while W-PGNN in \citet{liu_physics-guided_2024} adapts a residual structure, with output and state defined as sums of the physical and NN models. 
 Unlike \cite{schon2022multi},  which assumes that the states and their time derivatives are known, both W-PGNN and our proposed approach learn a mapping from input signal to the output measurement. 

One of the challenges for using these models that feed back the state into the NN for multi-step predictions, like W-PGNN, is that using non-saturating activations such as ReLUs can cause the state to increase exponentially. However, saturating activations, such as the sigmoid function used in \citet{relan2017unstructured} and the tanh used in \citet{schon2022multi}, suffer from vanishing gradients \citep{glorot2011deep}. Furthermore, the radial basis function used in W-PGNN only activate values close to the centers of the basis functions. These centers are parameters that need to be carefully tuned for each specific dataset. 
A separate challenge lies in bounding NN output, given the lack of knowledge of the NN component. These issues constrain the performance of physics-guided SSNNs, particularly when modeling a broader range of complex systems.

To address this issue, we propose a model that has an RNN layer in addition to the 2-layer fully connected (FC) feedforward networks used in W-PGNN to learn the temporal dependency in dynamics without feeding back states to the NNs.   This allows for safe use of ReLU activations to mitigate vanishing gradients and data-specific parameter tuning, without the risk of training instability and without the need to impose hard bounds on the state estimates.  
To allow for a high-dimensional hidden layer in the RNN and for more depth in the NN part to learn highly nonlinear functions, we use feature extraction layers to map the lower-dimensional inputs to higher dimensions inspired by \citet{gedon_deep_2021}.

\section{Method}

\label{sec:method}

We propose the PG-RSSNN to represent a SSM which is partially known from physics-based knowledge while the residual is modeled using a neural network as 

\begin{subequations} \label{eq:pa}
\begin{align}
x_t &= f_\text{phy} (x_{t-1}, u_t) + f_{\text{nn},\eta} (u_t), \label{eq:pa1a} \\
y_t &= g_\text{phy} (x_t, u_t) + g_{\text{nn},\sigma} (x_t, u_t). \label{eq:pa1b}
\end{align}
\end{subequations}

Here, $f_\text{phy}(\cdot)$ and $g_\text{phy}(\cdot)$ represent the known physical model. Furthermore, $ f_{\text{nn},\eta}(\cdot)$ and $ g_{\text{nn},\sigma}(\cdot)$ represent the unknown residual in the dynamic and measurement model, respectively. Note that $f_{\text{nn},\eta}(.)$ does not depend on previous state ${x}_{t-1}$. Instead, we adopt an RNN-based structure to capture temporal dependencies in $f_{\text{nn},\eta}(\cdot)$. Specifically, we select Gated Recurrent Units (GRU) \citep{chung_empirical_2014}, a widely used RNN variant known for its structural simplicity.  For notational simplicity, the explicit dependence of $f_{\text{nn},\eta}(\cdot)$ on $h_{t-1}$ is omitted in \eqref{eq:pa}. Note that \eqref{eq:pa} becomes a standard physics-only SSM if $ f_{\text{nn},\eta}(\cdot)$ and $ g_{\text{nn},\sigma}(\cdot)$  would be removed. Reversely, if no knowledge would be available such that $ f_{\text{phy}}(\cdot)$ and $ g_{\text{phy}}(\cdot)$ are zero, \eqref{eq:pa} becomes simply a black-box NN. 
The graphical structure for the proposed PG-RSSNN can be found in Fig.~\ref{fig:PG-RSSNN}. It implements \eqref{eq:pa} and additionally includes feature extraction and output layers for both $f_{\text{nn},\eta}(\cdot)$ and $g_{\text{nn},\sigma}(\cdot)$. 
 The feature extraction layers $\phi(\cdot)$ map low-dimensional inputs ($u_t \in \mathbb{R}^{n_\text{u}}$ and $\hat{x}_t \in \mathbb{R}^{n_\text{x}}$) to high-dimensional features ($\phi(u_t)$ and $\phi(\hat{x}_t)$) of size $n_\text{h}$, where $n_\text{h} > n_\text{u}, n_\text{x}$. The output layers subsequently map the learned high-dimensional features in the dynamic and measurement model back to the low-dimensional outputs $\hat{y}_t \in \mathbb{R}^{n_\text{y}}$.
The dynamic output layers, measurement output layers, and feature extraction layers are  implemented as two-layer FC networks, each with hidden layers of size $n_\text{h}$ and a ReLU activation between the layers.

We consider multi-step predictive learning: given $u_{1:T}$, we obtain
$\hat{y}_{1:T}$ by setting $\hat{x}_0 = 0$, propagating recursively through
\eqref{eq:pa1a}, and computing $\hat{y}_t$ via \eqref{eq:pa1b}. As $x_{1:T}$ is
unavailable, the model is trained on output loss only. Hence, $\hat{x}_t$ is not guaranteed to match the true physical state. However, $f_\text{phy}(\cdot)$ and $g_\text{phy}(\cdot)$ reduce what $f_{\text{nn},\eta}(\cdot)$ and $g_{\text{nn},\sigma}(\cdot)$ need to learn from data, improving data efficiency.

Unlike \citet{gedon_deep_2021}, we match the output dimension of $f_\text{nn}(\cdot)$ to the state dimension $n_\text{x}$, enabling direct addition with physics-based states $f_\text{phy}(\cdot)$. As shown in Fig~\ref{fig:PG-RSSNN}, this sum, denoted by $\hat{x}_t$, is fed into the measurement model. Similar to the dynamic model, the PG-RSSNN output $\hat{y}_t$ is the sum of NN output  $g_{\text{nn},\sigma} (\cdot)$ and physical output $g_\text{phy} (\cdot)$, both of dimension $n_\text{y}$.

The PG-RSSNN model is trained using the Adam optimizer \citep{kingma2014adam} with a learning rate of $10^{-2}$, to ensure fast and stable convergence. The training objective is 
\begin{equation}
\label{loss}
\min_{\eta,\sigma} \sum_{t=1}^T|| y_t-\hat{y}_t||^2_2.
\end{equation}


\section{Datasets and physical models}
\label{sec:dataset}

We study the workings of our algorithm on four different systems: a simulated linear Gaussian state-space model (LGSSM),  a simulated robotic arm, a real-world robotic arm \citep{weigand_dataset_nodate}, and a cascaded watertank system \citep{schoukens2020cascaded}. The simulated robotic arm dataset is generated based on the real robotic arm dataset, allowing controlled experiments to study the effect of different imperfections in the physical model.

\subsection{Linear Gaussian state-space model}
\label{sec:lgssm}
The first system we use is the LGSSM
\begin{equation}
\begin{aligned} 
       &    x_{t}= \overbrace{\begin{bmatrix} 0.7 & 0.8 \\ 0 & 0.1 \end{bmatrix}}^{A} x_{t-1}+\overbrace{\begin{bmatrix} -1 \\ 0.1  \end{bmatrix}}^{B} u_t + w_t, \\
       &    y_t=\underbrace{\begin{bmatrix} 1 & 0  \end{bmatrix}}_{C} x_{t} + v_t,
\end{aligned}
\label{eq:lgssm}
\end{equation}
with $w_t \sim \mathcal{N} \left( 0_2, 0.5 \, \mathcal{I}_2 \right)$ and $v_t \sim \mathcal{N} \left( 0, 1 \right)$. We trained and validated models on 50 datasets generated with different noise realizations $w_t$ and $v_t$, and evaluated these models on a single test set. Training and validation sequences have length 2000, and the test set sequence has length 5000. Given the system~\eqref{eq:lgssm}, we have $n_\text{u}=n_\text{y}=1, n_\text{x}=2$ in~\eqref{eq:pa}, and we choose $n_\text{h}=16$.





\subsection{Real world robotic arm dataset}
\label{sec:RealRobo}
We also consider a more complex nonlinear system with real-world data, the 6-joint KUKA robotic arm dataset \citep{weigand_dataset_nodate}. The dataset includes 36 trajectories, each having a duration of 60.6 seconds, performed twice (72 sequences in total), sampled at 10 Hz. We train the models to learn the mapping from the input joint torques $u_{1:\text{T}} \in\mathbb{R}^6$ to the output positions  $y_{1:\text{T}}\in\mathbb{R}^6$ in radians.
We write the torque-to-position system as the continuous-time state-space model
\begin{equation}\label{eq:EOM}
\begin{aligned}
{\dot{x}^\text{}} 
 =&{\begin{bmatrix}
{\dot{q}}\\
{\ddot{q}}
\end{bmatrix}}
 ={\begin{bmatrix}
{\dot{q}}\\
M(q) ^{-1}(u-C(q, \dot{q}) \dot{q} - g(q))
\end{bmatrix}}\\
y= &\begin{bmatrix}\mathcal{I}_{6} 
\quad 0_{6 \times 6}\end{bmatrix}x^\text{},
\end{aligned},
\end{equation}

 where $q$, $\dot{q}$ and $\ddot{q}$ represent continuous-time joint position, velocity and acceleration, respectively. The dynamics are given by the well-known robot model \citep{walker_efficient_1982}  with $M(q)$ the pose-dependent inertia matrix, $C(q,\dot{q})$ the Coriolis matrix and $g(q)$ the gravitational torques, where we use the physical knowledge of the 6-joint KUKA robot from \citep{weigand_dataset_nodate}.
To use the model \eqref{eq:EOM} in the PG-RSSNN  \eqref{eq:pa}, we first compute $\ddot{q}$ at a specific time instance $t$ and then use Runge-Kutta (RK4) numerical integration to obtain $q_{t+1}$ and $\dot{q}_{t+1}$.

We select the first 30 trajectories as the training set, trajectories 31--33 as the validation set, and trajectories 34--36 as the test set.\footnote{The last sequence contains fewer than 606 samples, so we pad it to 606 samples. The joint torques are padded using zeros and the positions are padded using the end pose. } Based on the number of robot joints and the state definition, we have $n_\text{u}=n_\text{y}=6, n_\text{x}=12$ in~\eqref{eq:pa}. Due to the complexity of the robot dynamic system, we choose $n_\text{h}=64$. Ten models are trained, each of which uses a different random seed to initialize the weights.

\subsection{Simulated robotic arm dataset}
\label{sec:simrobo}
To study the PG-RSSNN's performance for the higher-dimensional and nonlinear system from Section~\ref{sec:RealRobo} under more controlled conditions, we perform a numerical experiment simulating a new set of joint positions from the same KUKA robot model based on original trajectories in the real robotic dataset. This simulated dataset allows us to introduce artificial disturbances and biases. 

First,  we add a  linear disturbance term to the dynamic model in \eqref{eq:EOM}, representing for instance unmodeled friction, hydraulic torque, or inaccuracy in the robot configuration. Second, we consider that the measured input torques and joint positions are subject to additive, potentially non-zero mean noise, denoted as $w_t$ and $v_t$, respectively. This results in the updated model
\begin{equation}\label{eq:ssm_sim}
\begin{aligned}
{x}_{t} =&
f^\text{EOM}({x}_{t-1},u_{t}, w_t)+ \begin{bmatrix} 0_{6 \times 6} & 0_{6 \times 6} \\ 0_{6 \times 6} & S \end{bmatrix}x_{t-1},\\
\vspace{10pt}
{y}_{t} =&\begin{bmatrix}
\mathcal{I}_{6} & 0_{6 \times 6}
\end{bmatrix}{x}_{t}+v_t.\\
\end{aligned}
\end{equation}
 
The function $f^\text{EOM}(\cdot)$ represents the physics-based, discrete-time dynamic model, including the RK4 integration, see also Section~\ref{sec:RealRobo}. We sample $w_{t,i} \sim \mathcal{N}(\theta_i,0.1 \, \mathcal{I}_6)$, $\theta_i = [4,5,5,0.5,2,0.5]^\Transp$, for joint number $i=1,2,\hdots,6$, and measurement noise $v_t \sim \mathcal{N}(0_{6},0.1 \, \mathcal{I}_6)$. 
The mean of the process noise is specifically chosen to not be zero-mean and to differ per joint to model different torque measurement error magnitudes. Furthermore, we choose $ S=\text{diag}\begin{bmatrix}
    -0.05,0.1,-0.2,0.05,0.1,-0.1
\end{bmatrix}$ to represent a relatively small linear disturbance, which affects each joint differently. Note that we tested with different values of similar magnitude and got comparable results. Because of this, we only report the results for this specific case. 
Inspired by existing work on the robotic arm dataset \citep{weigand_dataset_nodate}, we take into account that the joint positions and velocities have physical constraints due to e.g.\ limited motion ranges of the KUKA robot. To this end, we apply a saturation on the output of $f^\text{EOM}$ both in the simulation as well as in the physics-guided networks. 


Using \eqref{eq:ssm_sim}, we generate and train models on 10 simulations with random $w_t$ and $v_t$, each containing the movements of all 72 sequences in \citet{weigand_dataset_nodate}. The input sequences are generated from acceleration, velocity and position using the Robotic toolbox \citep{corke_not_2021}. This implies that the simulation trajectories, as described in \eqref{eq:ssm_sim}, are similar to the real dataset but differ due to different noise realizations. The training and test set division as well as $n_\text{h}$ is the same as in Section~\ref{sec:RealRobo}.  



\subsection{Cascaded tank dataset}
\label{sec:realcascade}
We also consider the publicly available cascaded tank dataset \citep{schoukens2020cascaded}. We use the original test sequence containing 1024 samples as the test set, select the last 256 samples in the original training sequence as the validation set, and use the rest (768 samples) as the training set. 
 The physical system of this dataset is 
\begin{equation}\label{eq:castank}
\begin{aligned}
{x}_{t+1} &= {x}_t + T_\text{s} \left( A {x}_t + B u_t + {f}_{\text{nonlinear}}({x}_t) \right) + {w}_{t},\\
{y}_{t} &= {x_{2,t}} + {v}_{t},
\end{aligned}   
\end{equation}
where
\(
{x}_t = 
\begin{bmatrix} x_{1,t} \\ x_{2,t} \end{bmatrix},\quad
A = 
\begin{bmatrix} k_2 & 0 \\ -k_2 & k_5 \end{bmatrix},\quad
B = 
\begin{bmatrix} k_3 \\ 0 \end{bmatrix},\\
{f}_{\text{nonlinear}}({x}_t) =
\begin{bmatrix} - k_1 \sqrt{x_{1,t}} \\ k_1 \sqrt{x_{1,t}} - k_4 \sqrt{x_{2,t}} \end{bmatrix}
\)
 and $T_\text{s}=4$s. The model parameters in~\eqref{eq:castank} are identified and set to $k_1 = 0.0464$, $k_2 = 0.0003$, $k_3 = 0.0412$, $k_4 = 0.0586$, and $k_5 = 0.0039$. Additionally, $n_\text{u}=n_\text{y}=1$, and $n_\text{x}=2$ in~\eqref{eq:pa}. We simplify the model by assuming that the first tank does not overflow. This approximation introduces nonlinear, non-Gaussian modeling errors, which are denoted by  ${w}_t$ and  ${v}_t$ in \eqref{eq:castank}. We train each model with $n_\text{h} = 16$ hidden units, using early stopping with a patience of 100 epochs, and repeat the training 10 times with different random weight initializations in the NNs.

\section{Results}

\label{sec:result}
In this section, we study different properties of our proposed model on the four datasets from Section~\ref{sec:dataset}. We report accuracy in terms of the averaged root mean square error (RMSE) for the LGSSM and the cascaded tank dataset, and normalized RMSE (NRMSE), which is defined as the RMSE divided by the standard deviation of the ground truth, for the robotic datasets to normalize errors relative to the different variability of each joint. Besides (N)RMSE, we include the Correlation Coefficient (CC) between the predictions and the ground truth to quantify how well the prediction follows the trend of the ground truth temporal patterns (1 = perfect correlation, 0 = no correlation).
We also study training instability by reporting the number of failed runs, in which training terminated due to predicted states exceeding the floating-point range.

\subsection{Accuracy and training instability}
\label{sec:result-rnn}
To study the performance of the proposed PG-RSSNN and its capability to avoid training instability, we compare to nine other models:
\begin{itemize}
\item A model with the same structure as PG-RSSNN using sigmoid activation functions instead of ReLUs, referred to as PG-RSSNN-sgm, with sgm for sigmoid.
\item A model with the same structure as PG-RSSNN which additionally feeds back the state $\hat{x}_t$ to $f_\text{nn}(.)$, referred to as PG-RSSNN(fb), with fb for feedback.
\item A variant of PG-RSSNN in which the GRU is replaced by two FC layers, feeding states back to $f_\text{nn}(.)$ like \citet{liu_physics-guided_2024}, referred to as PG-FCSSNN.
\item  A model with the same structure as PG-FCSSNN, uses sigmoid activation, referred to as PG-FCSSNN-sgm. 
\item The corresponding black-box models of  PG-RSSNN and four above-mentioned physics-guided models, referred to as RSSNN,  RSSNN-sgm, RSSNN(fb), FCSSNN, FCSSNN-sgm, respectively.
\end{itemize}

Imperfect models are used in the physics-guided models, and training performance is assessed on learning to compensate for these imperfections. For the simulated robotic arm, \eqref{eq:ssm_sim} is used, knowing the disturbance while  $w_{t}$ and  $v_{t}$ are unknown. The models for the real robotic arm \eqref{eq:EOM} and for the cascaded tank system \eqref{eq:castank} are already imperfect due to inaccuracies in the robot definitions, sensor noises, and overflow of the first tank. Furthermore, for the LGSSM from Section~\ref{sec:lgssm}, slightly different system matrices $A=\begin{bmatrix} 0.6 & 0.7 \\ -0.1 & 1 \end{bmatrix}$,  $B=\begin{bmatrix} -1.1 \\ 0  \end{bmatrix}$ and $C=\begin{bmatrix} 0.9 & -0.1  \end{bmatrix}$ are used ascompared to \eqref{eq:lgssm}. 
The five physics-guided models we compare all use the same physical model.

\begin{table}[htbp]
\centering
\caption{
Average (N)RMSE and training success rate in parentheses (only if runs failed),  lowest (N)RMSE per dataset in bold.  \label{tab:goodstructure}
}
\renewcommand{\arraystretch}{1.2}
\begin{tabular}{l@{\hspace{3pt}}c@{\hspace{3pt}}@{\hspace{3pt}}c@{\hspace{3pt}}@{\hspace{3pt}}c@{\hspace{3pt}}@{\hspace{3pt}}c@{\hspace{3pt}}}

\Xhline{2\arrayrulewidth}
 & LGSSM & \makecell{Real\\Robotic} & \makecell{Simulated\\Robotic} & \makecell{Cascaded\\Tank} \\
\Xhline{2\arrayrulewidth}
FCSSNN-sgm & 1.88 &0.99& 0.94& 2.11\\
PG-FCSSNN-sgm &1.38  &1.23&0.87 &0.73 \\
RSSNN-sgm & 1.88 & 1.00  &0.91  & 1.28 \\ 
PG-RSSNN-sgm & 1.40 & 1.23  & 0.74  & 0.55 \\ 
\hline
FCSSNN & 1.39 & {0.95 ({9/10})} & 0.90  & {1.70 ({6/10})} \\
PG-FCSSNN & \textbf{1.37} & {0.89 (1/10) }& 0.65  & 0.71  \\
RSSNN(fb) & 1.40 & {0.94 ({9/10})} & {0.91 ({5/10})} & 1.49  \\
PG-RSSNN(fb) & \textbf{1.37} & {- (0/10)} & {0.67 (6/10)} & 0.46 \\
\hline
RSSNN & 1.39 & 0.96  & 0.92  & 0.71 \\

\textbf{PG-RSSNN}& 1.38 & \textbf{0.86} & \textbf{0.53}  & \textbf{0.44}  \\
\Xhline{2\arrayrulewidth}
\end{tabular}
\end{table}

Table~\ref{tab:goodstructure} reports the RMSEs for LGSSM and the cascaded tank dataset, and NRMSEs for the robotic datasets.
The models with ReLUs that do not feed back the state (PG-RSSNN, RSSNN) and the models using sigmoid functions (PG-RSSNN-sgm,  RSSNN-sgm, PG-FCSSNN-sgm, and FCSSNN-sgm) consistently avoid training instability, while training instability does occur for the models that do feed back the states and also use ReLUs (FCSSNN,PG-FCSSNN,RSSNN(fb) and PG-RSSNN(fb)). This seems to occur more often when the models are higher dimensional and more nonlinear, since no instability was observed for the LGSSM, while PG-RSSNN(fb) fails 100\%  on the real robotic dataset. 

 In terms of prediction accuracy, the proposed model PG-RSSNN achieves the lowest (N)RMSE on three of the four datasets, and comparable performance on the remaining LGSSM dataset, with an RMSE only 1\% higher than the best model. 
 In contrast, its sigmoid-based counterpart, PG-RSSNN-sgm, matches PG-RSSNN performance for the LGSSM, but shows higher errors on the remaining datasets. This trend holds across other model pairs: ReLU-based models (FCSSNN, PG-FCSSNN, RSSNN, PG-RSSNN) consistently outperform their SGM variants (FCSSNN-sgm, PG-FCSSNN-sgm, RSSNN-sgm, PG-RSSNN-sgm). Although PG-FCSSNN-sgm preserves training stability and performs similarly on the LGSSM, it exhibits 43\%, 39\%, and 66\% higher (N)RMSE than PG-RSSNN on the real and simulated robotic datasets and on the cascaded tank dataset, highlighting that ReLU and RNN’s benefits  becomes more apparent for nonlinear systems. Overall, these results indicate that, in our experiments, the improved predictive performance of PG-RSSNN can be attributed to the use of ReLU activation functions.
 
Studying the effect of using physical knowledge, it can be seen that the physics-guided models using ReLU activation functions (PG-RSSNN, PG-RSSNN(fb) and PG-FCSSNN) have a lower (N)RMSE compared to their black-box counterparts (RSSNN, RSSNN(fb) and FCSSNN, respectively). 
Compared with a state-feedback feedforward model of similar complexity (PG-FCSSNN), PG-RSSNN achieves comparable performance on the LGSSM and real robotic dataset, with differences under 3\%.
However, on the other two datasets, PG-RSSNN performs substantially better than PG-FCSSNN by 19\% and 38\%, respectively. 

Table \ref{tab:goodstructure} shows that PG-RSSNN(fb) and RSSNN(fb) do not outperform their non-feedback counterparts (PG-RSSNN and RSSNN) in terms of multi-step prediction accuracy. On the simulated robotic dataset, the NRMSE of PG-RSSNN(fb) (0.67) is even 26\% higher than that of PG-RSSNN (0.53), suggesting that including state feedback in addition to RNN structure in PG-RSSNN does not improve prediction accuracy, and even results in occasional training instability. Thus, we conclude that the GRU layer in PG-RSSNN effectively replaces the state-feedback structure used in non-RNN structure.
\subsection{Different imperfect physical models}
\label{sec:result-physics}
This section evaluates PG-RSSNN's performance under different types of imperfect physical models and demonstrates its workings over the black-box and physics-only models. 

Besides the imperfect physical models described in Section~\ref{sec:result-rnn}, referred to as {Model~1}, we additionally evaluate two alternative physical models for the two simulated datasets to assess prediction accuracy under different levels of misspecification. For the LGSSM dataset, {Model~2} assumes the measurement model to be unknown while treating the state-transition matrix $A$ as correctly specified and $B$ as unknown, i.e., $f_{\text{phy}}(\cdot)=A x_{t-1}$ and $g_{\text{phy}}(\cdot)=0$. {Model~3} instead assumes $A = \begin{bmatrix} 0.6 & 0.7 \\ -0.1 & 1 \end{bmatrix}$, and $ B = C = 0$. For the simulated robotic arm system, {Model~2}  assumes the linear disturbance is unknown corresponding to using the dynamics and measurement model in~\eqref{eq:EOM}, whereas {Model~3} assumes that only the physical dynamics $f^{\text{EOM}}$ from~\eqref{eq:EOM} are known, i.e.\ $f_{\text{phy}}(\cdot)=f^{\text{EOM}}$ and $g_{\text{phy}}(\cdot)=0$.






\begin{table}[h!]
\centering
\caption{ (N)RMSE for different imperfect physical models, lowest (N)RMSE per dataset in bold.}
\label{tab:physical_model_comparison}

\begin{tabular}{l l c c c}
\hline
\makecell{Dataset} &
\makecell{Physical\\model} &
\makecell{Physics-\\only} &
RSSNN &
\makecell{PG-\\RSSNN} \\
\hline

\multirow{3}{*}{LGSSM} 
    & Model 1 & 1.68 & \multirow{3}{*}{1.39} & \textbf{1.38} \\
    & Model 2 & --   &                       & \textbf{1.38} \\
    & Model 3 & --   &                       & \textbf{1.38} \\
\hline

\multirow{3}{*}{\makecell{Simulated\\robotic arm}}
    & Model 1 & 0.54 & \multirow{3}{*}{0.92} & \textbf{0.53} \\
    & Model 2 & 0.81 &                       & \textbf{0.71} \\
    & Model 3 & --   &                       & \textbf{0.67} \\
\hline
\end{tabular}
\end{table}

In Table~\ref{tab:physical_model_comparison}, the performance of different experiments and models over 2 datasets is demonstrated. It can be seen that the proposed PG-RSSNN outperforms both the RSSNN and the physics-only model, irrespective of the physical models that are used. Note that the RSSNN does not use any physical model, so only one (N)RMSE per dataset is reported. The dashes indicate physics-only simulation is unavailable for those physical models without the knowledge of the measurement model.
Notably, the larger improvement on the simulated robotic dataset (41.9\% vs.\ 10.7\% on the real robotic dataset) is expected, as the physical model is more accurate there (NRMSE 0.54 vs.\ 3.33).
It is not surprising that the improvement of using PG-RSSNN is small when rather inaccurate physical knowledge is provided.

\subsection{Limited real-world training data}
\label{sec:result-lessdata}

We also evaluate the performance of the proposed PG-RSSNN for limited training data. To this end, we use the cascaded tank dataset, which in its entirety is already considered small~\citep{schoukens2020cascaded}, and evaluate PG-RSSNN trained on subsets of
\(16.7\%\), \(33.3\%\), \(50\%\), and \(100\%\) of the training data, corresponding to \(128\), \(256\), \(384\), and \(768\) samples, respectively (i.e., 1, 2, 3 and 6 times the base training sequence length of \(128\)).
\begin{figure}[h]
\centering
\begin{subfigure}[b]{0.46\linewidth}  
    \centering
\hspace{-0.55cm}
    \includegraphics[width=1.1\linewidth, trim=0cm 0cm 0cm 0cm, clip]{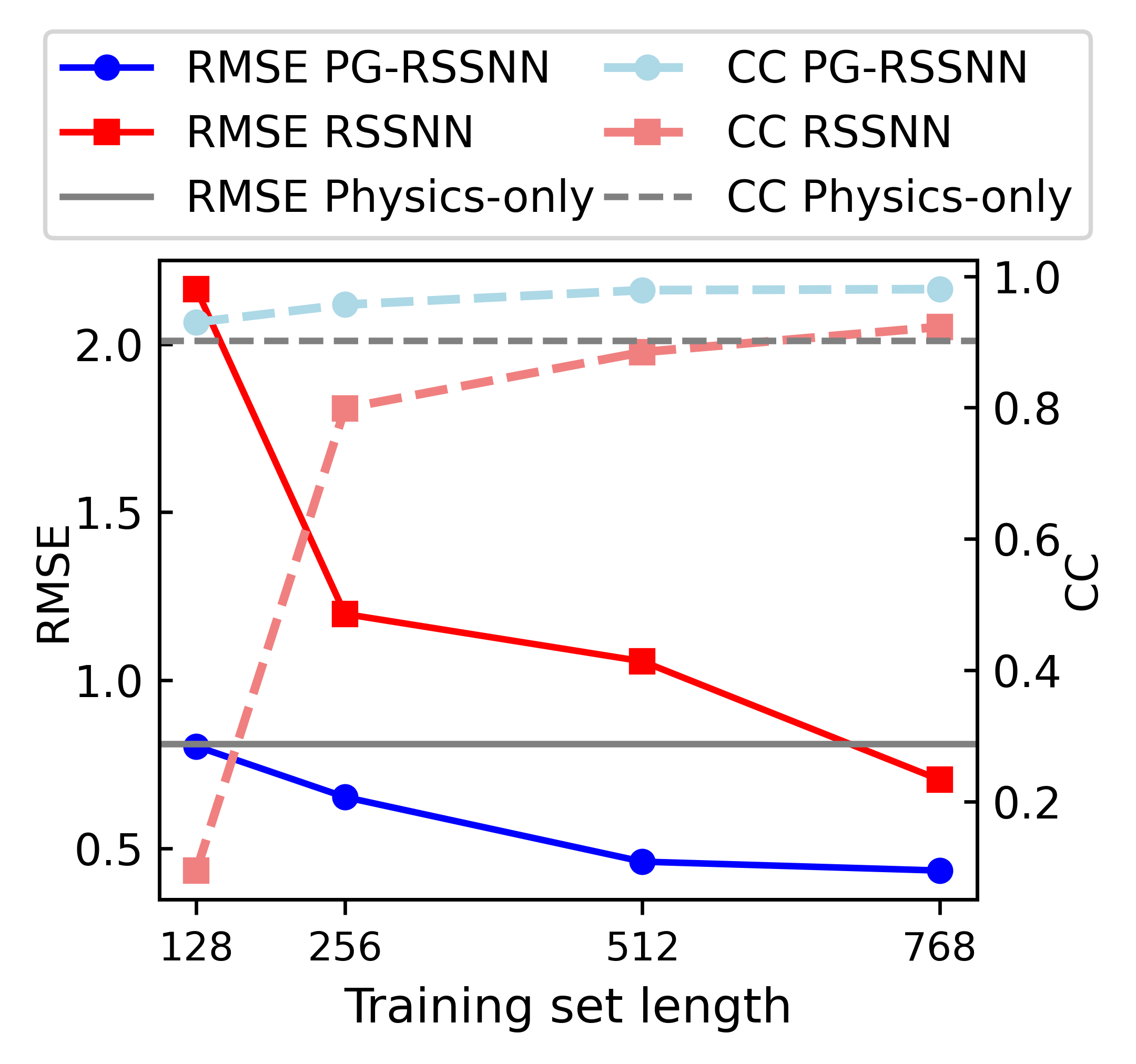}
    \caption{RMSE and CC of \\different models.}
    \label{fig:2datasetdatasetsize}
\end{subfigure}
  \hspace{-0.5cm}
\begin{subfigure}[b]{0.52\linewidth}  
    \centering
    \includegraphics[width=1.1\linewidth, trim=0.2cm 0.9cm 0cm 1.3cm, clip]{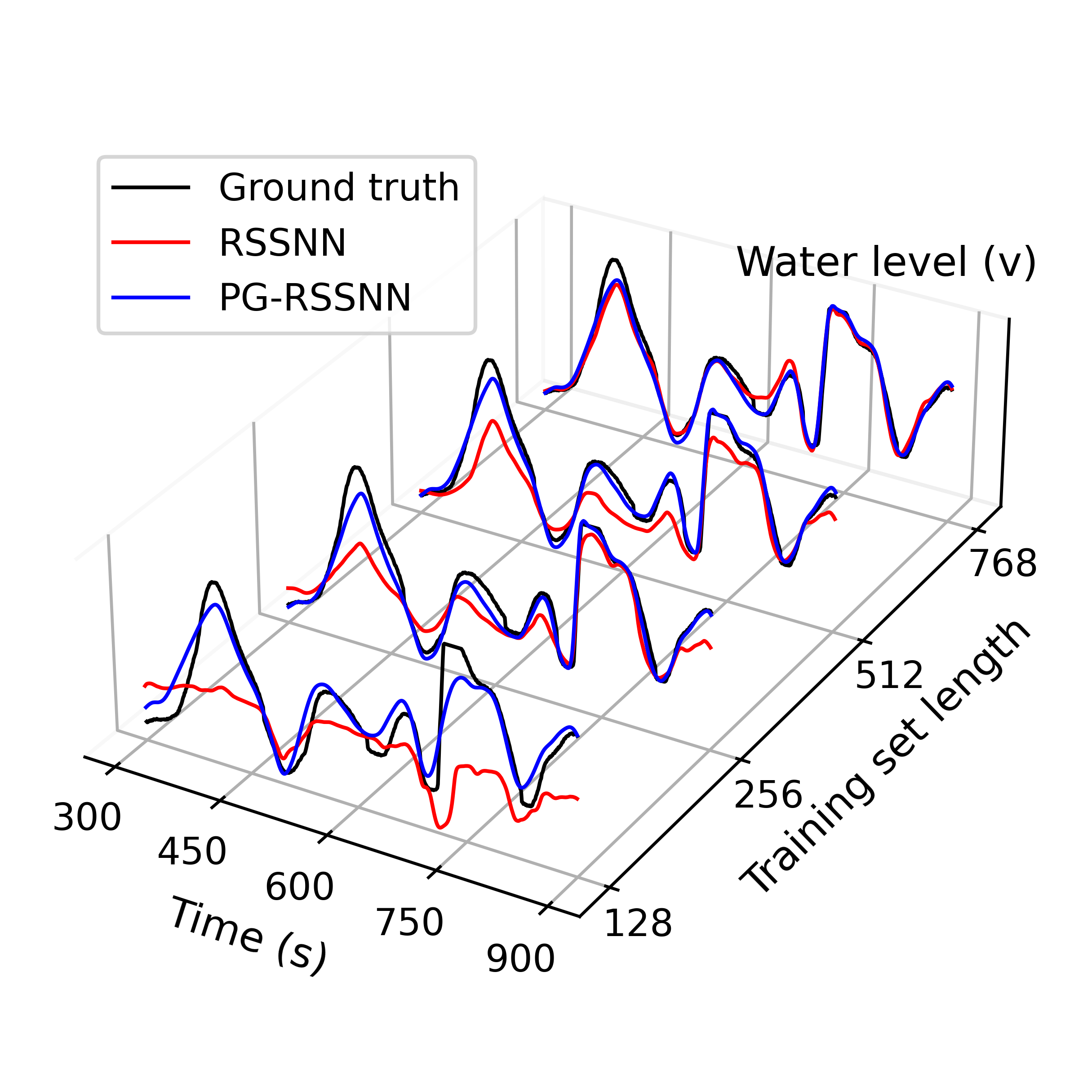}
    \caption{Predicted and ground truth\\ water level in volt (V).}
    \label{fig:casestprediction}
\end{subfigure}
\caption{Cascaded tank dataset results for different training set lengths. (a) RMSE and CC of PG-RSSNN, RSSNN, and physics-only model. (b) Predictions of PG-RSSNN and RSSNN compared with ground truth.}
\label{fig:casc_tank_subplots}
\end{figure}

As shown in Fig~\ref{fig:2datasetdatasetsize}, when the dataset length decreases from 100\% to 50\%, the CC of the RSSNN drops below that of the physical model, indicating a rapid decline in the RSSNN's ability to capture temporal patterns. In contrast, even when trained on only 17\% of the full dataset size, the PG-RSSNN achieves a higher CC and lower RMSE than both the physical model and the RSSNN. Across the four dataset sizes, the RMSEs of PG-RSSNN remain consistently lower than those of the RSSNN.
Fig.~\ref{fig:casestprediction} shows the rapid degradation of the black-box RSSNN with less training data. Even when the training set is reduced to 128 samples, PG-RSSNN (blue curve) still closely follows the ground-truth water level (black curve), while the RSSNN (red curve) prediction significantly deviates. This illustrates the PG-RSSNN's effectiveness in predicting the system's behavior, even when training datasets are small.

Note that the average and best performances of PG-RSSNN (average RMSE = 0.44, best RMSE = 0.34) trained on 100\% training set, and  the model with only 50\% of the original training set (average RMSE = 0.46, best RMSE = 0.36), are comparable to state-of-the-art methods that use the entire cascaded tank dataset for training. These include NLSS2 (RMSE = 0.34) \citep{relan2017unstructured} and the black-box GRU model (RMSE = 0.40) \citep{champneys2024baseline}. 

\section{Conclusion and Future Work}
\label{sec:conclusion}
We propose PG-RSSNN, an RNN-based physics-guided SSNN for multi-step prediction.  Unlike traditional SSNNs, it learns temporal dependencies via an RNN structure without feeding the estimated state back. We have shown that PG-RSSNN safely uses the non-saturating ReLU activation function to mitigate vanishing gradients and system-specific tuning. The proposed model can straightforwardly be used for different systems by adjusting only the input, state, output, and hidden layer dimensions. It exhibits no training instability in any of the real-world and simulated datasets considered, and outperforms black-box NNs, several variants of physics-guided NNs and physics-only models, even with imperfect physical models, for the considered datasets.
Furthermore, on a subset of the cascaded tank dataset with only 128 samples, PG-RSSNN still improves over purely black-box or physics-only models, demonstrating its data efficiency. Our focus in this work was only to adapt the model structure to combine data-driven NNs with known physical models and avoid training instability. In future work, it is possible to adapt the loss function, e.g.\ incorporating a physics-based penalty to improve the physical plausibility of predictions  \citep{daw_physics-guided_2021}. 
\begin{ack}
The authors gratefully acknowledge Dr. Yuhan Liu for her code and parameters of the cascaded tank system and Dr. Jonas Weigand for his support on the technical details of the KUKA robotic arm model.
\end{ack}

\section*{DECLARATION OF GENERATIVE AI AND AI-ASSISTED TECHNOLOGIES IN THE WRITING PROCESS}
During the preparation of this work the author(s) used ChatGPT in order to check the grammar and improve writing. After using this tool/service, the author(s) reviewed and edited the content as needed and take(s) full responsibility for the content of the publication.

\bibliography{ifacconf}

\end{document}